

\magnification=1200

\vsize=7.5in
\hsize=5.5in
\tolerance 10000
\baselineskip 12pt plus 1pt minus 1pt
\pageno=0
\centerline{{\bf FOUR-PARAMETER POINT-INTERACTION IN 1-D QUANTUM SYSTEMS}
\footnote{*}{This work is supported in part by funds
provided by the U. S. Department of Energy (D.O.E.) under contract
\#DE-AC02-89ER40509, and by the Natural Science and Engineering
Research Council Of Canada.}}
\vskip 24pt
\centerline{Michel Carreau}
\vskip 12pt
\centerline{\it Boston University}
\centerline{\it Department of Physics}
\centerline{\it 590 Commonwealth Avenue}
\centerline{\it Boston, Massachusetts\ \ 02215\ \ \ U.S.A.}
\centerline{E-mail:CARREAU@BUPHY.BU.EDU}
\vskip 1.5in
\centerline{\bf ABSTRACT}
\medskip
We construct a four-parameter point-interaction for a
non-relativistic particle moving on a line as the limit of a short range
interaction with range tending toward zero.
For particular choices of the
parameters, we can obtain a $\delta$-interaction or the so-called
$\delta'$-interaction.
The Hamiltonian corresponding to the four-parameter
point-interaction is shown to correspond to the four-parameter
self-adjoint Hamiltonian of the free particle moving on
the line with the origin excluded.
\bigskip
\bigskip
\bigskip
\centerline{To be published in: {\it Journal of Physics} \bf A}
\vfill
\centerline{ Typeset in $\TeX$ }
\vskip -12pt
\noindent BUHEP-92-9\hfill December 1991
\eject
\noindent{\bf I.\quad INTRODUCTION}
\medskip
\nobreak

The study of point-interactions in
non-relativistic quantum mechanics has  been studied extensively
in recent years$^1$. The interest in this subject is two folds. First there is
the hope that a point-interaction can be a good approximation of a very
localized interactions. Second, It is possible to obtain exact solution
for  quantum systems with point-interactions.
Physically, a very localized interaction can be  due to the
interaction of a particle with an impurity or a local defect in a solid for
example. It could also be due to the region of contact(point-contact)
between two conducting materials,...,etc.

In two or three dimensions a  point-interaction can be
thought of as the interaction of a particle with a
$\delta$-potential with a renormalized coupling constant$^2$.
However it is also possible to describe the point-interaction in terms of
boundary conditions on the wave functions at the interaction point which is
excluded from the configuration space$^1$. In this case the different
strengths of the interaction is characterized by different boundary
conditions.


In one dimension, the situation is different.
Before our  paper, the most general point-interaction in one dimension could
only be expressed in terms of boundary conditions which we review below.
The object of our  paper is to construct a short-range interaction
which, in the zero-range limit, gives a physical realization of the
point-interaction in one dimension.

The general point-interaction in
one dimension is obtained by considering the self-adjoint-extensions
of the Hamiltonian of a free particle moving on a line with
the origin excluded (see Fig.~1). It is found that there is a
four-parameter family of self-adjoint Hamiltonians that can
be characterized  by a four-parameter family of boundary condition
imposed on the wave functions$^{3,4}$.  Let us recall this boundary
condition, following  the notation of Ref.~[4].
For the wave function $\psi(x)$,
defined everywhere except at $x=0$, we require
$$\left[ \matrix{ -\psi'_L \cr\noalign{\vskip 0.2cm} \psi'_R \cr}\right] = M
\left[\matrix{ \psi_L\cr\noalign{\vskip 0.2cm} \psi_R\cr} \right] \eqno(1.1)$$
where
$$\eqalign{\psi_R &\equiv \lim\limits_{\epsilon\to 0^+}
\psi(\epsilon)\cr\noalign{\vskip 0.2cm} \psi_L &\equiv
\lim\limits_{\epsilon\to 0^-} \psi(\epsilon) \cr} \hskip .5in
\eqalign{ \psi'_R &\equiv \lim\limits_{\epsilon\to 0^+} {d\psi\over dx}
(\epsilon) \cr\noalign{\vskip 0.2cm}
\psi'_L &\equiv \lim\limits_{\epsilon\to 0^-} {d\psi\over dx} (\epsilon) \cr}
\eqno(1.2)$$
and $M$ is an arbitrary $2\times 2$ Hermitian matrix
which  can be parametrized in terms
of four real parameters
$$M = \left(\matrix{ \rho+\alpha & -\rho\,e^{i\theta} \cr\noalign{\vskip
0.2cm} -\rho\,e^{-i\theta} & \rho + \beta \cr}\right)\eqno(1.3)$$
with $\rho\ge 0$ and $0\le\theta<2\pi$.
If the domain of the Hamiltonian consists of wave functions
obeying (1.1) for a fixed $M$, then the Hamiltonian, will be self-adjoint
and will be denoted $H_M$.
This boundary condition ensures the conservation of probability at
the origin or, equivalently, that the current of probability is
continuous through the origin.

Note that, for the quantum system consisting
of a free particle restricted to move inside an interval of length $L$,
we have a similar story. We can imagine bringing the
extremities of the interval close to each other,  making it
looking like a circle with a hole, see fig.~1. The result of our
paper apply to this system as well.

An interpretation of the parameters $\rho,\alpha,\beta,\theta$
in the context of functional integral was given in Ref. [4].
It was found that the measure on paths in the functional integral
is controlled by these parameters.  Let us recall, also from Ref. [4],
that the current of probability at the origin is
proportional to $\rho$.  For $\rho=0$, the boundary condition
reduces to $-\psi'_L=\alpha\psi_L$ and $\psi'_R=\beta\psi_R$
which describes the physics of two separate half-lines. For
$\rho$ infinite, $\theta=0$, and, $\alpha$ and $\beta$ finite,
the wave function is  continuous, and $\psi'_R-\psi'_L=(\alpha+\beta)\psi_L$
which is the $\delta$-interaction. For $\rho$ finite and
$\theta=\alpha=\beta=0$, the derivative of the wave function is
continuous and $\psi_R-\psi_L={1\over\rho}\psi'_L$ which is the
so-called $\delta'$-interaction (see eg. Ref. [1]).

In this paper, we construct a Hamiltonian defined on the
whole line which described a four-parameter family point-interaction.
The interacting terms in the  Hamiltonian provides a physical understanding of
the four parameters.

The organization of the paper is as follows:
in Section II, we construct a four-parameter
point-interaction. We show that the effect of adding this
point-interaction to the free Hamiltonian on the whole line
leads to a system that is identical to the free particle
on the line with a hole with the boundary condition (1.1).
In Section III, we study the scaling properties of the
point-interaction.
\goodbreak
\bigskip
\noindent{\bf II.\quad POINT-INTERACTION}
\medskip
\nobreak
In this section, we construct a four-parameter point-interaction
for a non-relativistic particle moving on the whole line. We define the
point-interaction as the limit of a local interaction of range
of order $\epsilon$ with $\epsilon$ going to zero.  We
construct the local interaction in such a way that, in the limit
where $\epsilon$ goes to zero, it forces the wave function to
satisfy the boundary condition (1.1).  In particular, we will
see that the so-called $\delta'$-interaction is obtained as the
zero-range limit of an operator that is non-self-adjoint for
finite range (see Ref. [5] for an alternative representation of
the $\delta'$-interaction).

We would like to mention that progress in this direction by P.
\v Seba$^{3}$ led to a two-parameter point-interaction expressed
formally as a sum of terms involving $\delta(x)$ and $\delta'(x)$.
However, the physical interpretation of this formal expression
seems unclear to us.

Let us consider the following Hamiltonian for the local interaction
depending upon the parameters $\rho,\theta,\alpha,\beta$
$$H^\epsilon=-{1\over 2} \ {d^2\over dx^2} + I^\epsilon\eqno(2.1)$$
where
$$I^\epsilon=\left\{
\matrix{K^\epsilon\!(\rho\,e^{i\theta})
\left({d\over dx}+\alpha+\rho-\rho\,e^{i\theta}\right)&\quad
-\epsilon<x<0\cr\noalign{\vskip 0.2cm}
-K^\epsilon\!(\rho\,e^{-i\theta})\left({d\over dx}+
\beta+\rho-\rho\,e^{-i\theta}\right)&\quad
0<x<\epsilon\cr\noalign{\vskip 0.2cm}
0 \quad &\hbox{elsewhere\ }\cr}\right\}\eqno(2.2)$$
is a local interaction. Also, $K^{\epsilon}(\eta)$ is
the solution of the transcendental equation
$$K^{\epsilon}\!(\eta)e^{-2K^{\epsilon}\!(\eta)\epsilon}=\eta\eqno(2.3)$$
for which $K^{\epsilon}\!(\eta)$ goes to infinity as $\epsilon$
goes to zero. The solution can be
written as an expansion for small $\epsilon$, that is
$$K^{\epsilon}\!(\eta)={1\over 2\epsilon}\left\{-ln(2\eta\epsilon)
+ln\left[-ln(2\eta\epsilon)+ln\left[\ldots\right]\right]\right\}
\ \ .\eqno(2.4)$$

This Hamiltonian describes the motion of a particle which moves under the
influence of the interaction $I^\epsilon$ only whenever it is inside
the interval $[-\epsilon,\epsilon]$, otherwise it moves freely.

We observe that the
Hamiltonian $H^\epsilon$ is not self-adjoint (not even Hermitian)
since the coefficient of ${d\over dx}$ is not purely imaginary.
However, we will show that, in the limit where $\epsilon$ goes to
zero, $H^\epsilon$ becomes self-adjoint.  Specifically, we will
show that, given a particular value of the parameters
$\rho,\alpha,\beta$ and $\theta$,
$H^\epsilon$ converges toward the self-adjoint Hamiltonian $H_M$
described in the previous section for the particle on the line with the
origin excluded.

To show that, we simply demonstrate that in the limit where
$\epsilon$ goes to zero, the energy eigenvalues of $H^\epsilon$ and the
corresponding energy eigenstates are identical to those of
$H_M$. Let us solve the eigenvalue equation for the Hamiltonian $H^\epsilon$
$$H^\epsilon\psi_E=E\,\psi_E\ \ .\eqno(2.5)$$
This is easy, we just solve this equation for each
interval for which it is a constant-coefficient differential
equation, and matches the solutions at the junctions of the different
intervals such that
the wave function and its derivative are continuous for all $x$.
Once this is done, we simply let $\epsilon$ tends toward zero.
In the limit of small $\epsilon$, we find
$$\psi_E(x)\approx\left\{
\matrix{\psi_L\,\hbox{cos}[k(x+\epsilon)]+{\psi'_L\over
k}\,\hbox{sin}[k(x+\epsilon)] &\quad x\le -\epsilon\cr\noalign{\vskip 0.2cm}
\psi_{L}\,e^{-D_\alpha x}+{e^{-i\theta}\over 2\rho}
\left[\psi'_L+D_\alpha\psi_L\right]
e^{\left[2K^{\epsilon}\!\left(\rho\,e^{i\theta}\right)+D_\alpha\right]x}
&\quad -\epsilon<x<0\cr\noalign{\vskip 0.2cm}
\psi_{R}\,e^{\tilde D_\beta x}-{e^{i\theta}\over 2\rho}
\left[\psi'_R-\tilde D_\beta\psi_R\right]
e^{\left[-2K^{\epsilon}\!\left(\rho\,e^{-i\theta}\right)-
\tilde D_\beta\right]x}
&\quad 0<x<\epsilon\cr\noalign{\vskip 0.2cm}
\psi_R\,\hbox{cos}[k(x-\epsilon)]+{\psi'_R\over k}\,\hbox{sin}[k(x-\epsilon)]
&\quad x\ge\epsilon\cr}\right\}
\eqno(2.6)$$
where
$$D_\alpha=\alpha+\rho-\rho\,e^{i\theta}\eqno(2.7a)$$
$$\tilde D_\beta=\beta+\rho-\rho\,e^{-i\theta}\eqno(2.7b)$$
$$K^\epsilon\!(\rho\,e^{\pm i\theta})={1\over 2\epsilon}\left\{
-ln(2\rho\epsilon)\mp i\theta\right\}+\ldots\eqno(2.7c)$$
We have also set $\psi_{L}\equiv\psi_E(-\epsilon)$,
$\psi'_{L}\equiv{d\over dx}\psi_E(x)|_{x=-\epsilon}$,
$\psi_{R}\equiv\psi_E(\epsilon)$,
$\psi'_{R}\equiv{d\over dx}\psi_E(x)|_{x=\epsilon}$, $E\equiv {k^2\over
2}$ and
$$\left[ \matrix{ \psi'_R \cr\noalign{\vskip 0.2cm} \psi_R
\cr}\right]=e^{-i\theta}
\left[\matrix{1+{\beta\over\rho}&\alpha+\beta+{\alpha\beta\over\rho}
\cr\noalign{\vskip 0.2cm}
{1\over\rho} & 1+{\alpha\over\rho}\cr}\right]
\left[\matrix{ \psi'_L\cr\noalign{\vskip 0.2cm} \psi_L\cr}\right]\ \ .
\eqno(2.8)$$

We can now easily check that $\psi_E(x)$ is continuously
differentiable and that it satisfies eq. (2.5) to leading orders,
in the limit of small $\epsilon$. Since (2.8) is actually
identical to the boundary condition (1.1), we immediately see that
the energy eigenstate, $\psi_E(x)$, in the limit where $\epsilon$ goes to
zero, satisfies the boundary condition (1.1), for all $E\equiv
{k^2\over 2}$.
Moreover, the energy, $E$, must be real in order to have finite energy
eigenstates at $x$ equal plus or minus infinity.
Therefore, the energy eigenstates, $\psi_E$, with energy $E={k^2\over 2}$,
of $H^\epsilon$, in the
limit of zero $\epsilon$, are identical to those of the Hamiltonian $H_M$,
as we wanted to show. (a more delicate technique,  which is more abstract
though, to show the convergence of a sequence of operators which become
singular in the limit can be found in the paper of Albeverio and \v Seba,
Ref.~[2,3], and I urge the interested readers to read these papers).

Let us make few comments about the short range interaction $I^\epsilon$.
The parameter $\theta$ is responsible for the phase discontinuity
of the wave function has can be seen from (2.8).
The parameter $\rho$  control the size of the discontinuity of
the wave function. This can be seen explicitly by setting
$\theta=\beta=\alpha=0$ in (2.7). The boundary condition
becomes $\psi_R-\psi_L={1\over\rho}\psi'_L$ and $\psi'_R=\psi'_L$ which is the
so-called $\delta'$-interaction.  We observe that, in (2.6),
the derivative of the wave function at the origin,
$\psi'_E(x)|_{x=0}$, is proportional to $K^\epsilon\!(\rho\,e^{i\theta})$ and
becomes infinite in the limit where $\epsilon$ goes to zero which
forces the discontinuity of the wave function.

\goodbreak
\bigskip
\noindent{\bf III.\quad SCALING}
\medskip
\nobreak
In this section, we study the scaling properties of the
point-interaction described in the previous section.

Consider the following transformation
$$\eqalign{x &\rightarrow\lambda x\cr (\alpha,\beta,\rho)
&\rightarrow{1\over\lambda}
(\alpha,\beta,\rho)\cr \theta &\rightarrow\theta\cr}\ \ .\eqno(3.1)$$
We can readily see that, for the quantum system consisting of a
free particle moving on a line with the origin excluded, discussed
in the first section, the
four-parameter boundary condition, (1.1), is invariant
under the above transformation. Since, in the previous section, we
have shown that the point-interaction has the effect of forcing
dynamically the boundary condition (1.1) on the wave function, we expect
that the Schr\"odinger equation, at the location of the
point-interaction, is invariant
under the above transformation.

To see that, we recall that in the previous section, we have defined
the point-interaction as the limit
of the local interaction (2.2) for which $\epsilon$ goes to zero.
We can readily see that, under  the above transformation,
with $\epsilon$ going to $\lambda\epsilon$ also, the
Schr\"odinger equation
$$H^\epsilon\psi(x,t)={1\over i}{d\over dt}\psi(x,t)\ \ .\eqno(3.2)$$
would be invariant if the term ${1\over i}{d\over dt}\psi(x,t)$
were not present.
However, in the interval $[-\epsilon,\epsilon]$ and in the limit
where $\epsilon$  goes to zero,
the term ${1\over i}{d\over dt}\psi(x,t)$ becomes arbitrarily
small relative to the
other terms.  Therefore, in the limit where $\epsilon$ goes to
zero, we see that
the Schr\"odinger equation becomes invariant under the transformation (3.1)
in the interval $[-\epsilon,\epsilon]$.
Which is what we expected.

\goodbreak
\bigskip
\noindent{\bf IV.\quad CONCLUSIONS}
\medskip
\nobreak
We have constructed a quantum system for a particle moving on
the whole line, subject to a local interaction, and have shown that
in the limit where the range of the interaction tends to
zero (point-interaction) our
quantum system tends toward the quantum system of a free moving
particle on the line with the origin excluded. The main point in
the demonstration of this result was the observation that
the point-interaction dynamically forces the wave function to
satisfy a boundary condition which ensures conservation of
probability at the origin. We also observed that, at the location
of the point-interaction, the physics  is invariant under a certain
scale transformation.

\goodbreak
\bigskip
\centerline{\bf ACKNOWLEDGMENTS}
I thank Edward Farhi, Sam Gutmann and the referees for many valuable
comments.

\par
\vfill
\eject
\centerline{\bf REFERENCES}
\medskip
\item{1.}S. Albeverio et al., {\it Solvable Models in Quantum Mechanics\/}
(Springer-Verlag, New York, 1988) and references therein.
\medskip
\item{2.}S. Albeverio, J. E. Fenstad and R. Hoegh-Krohn{\it Trans. Am. Math.
Soc.} {\bf 252} (1979) 275.

R. Jackiw, {\it M.A.B. B\'eg Memorial Volume} , (A. Ali, P. Hoodbhoy, Eds)
World Scientific, Singapore, 1991; and MIT-CTP-1937
\medskip
\item{3.}P. \v Seba, {\it Czech. J. Phys.} {\bf 36} (1986) 667.
\medskip
\item{4.}M. Carreau, E. Farhi, S. Gutmann, {\it Phys. Rev.} {\bf D42}
(1990) 1194.
\medskip
\item{5.}P. \v Seba, {\it Rep. Math. Phys. Vol.} {\bf 24} (1986) 111.
\medskip
\goodbreak
\bigskip
\vfill
\eject
\noindent{\bf Figure Captions}
\medskip
\noindent
Fig. 1: The upper diagram shows the line with a hole.  The lower
diagram shows the box distorted to indicate the similarity
between the boundary conditions at the walls of the box and the
two sides of the hole.
\vfill
\end